\documentclass[twocolumn]{aastex631}

\usepackage{amsmath, amstext, verbatim, graphicx, xcolor, longtable, multirow}
\usepackage{placeins}

\newcommand{\Hb}{\hbox{{\rm H}$\beta$}}
\newcommand{\Ha}{\hbox{{\rm H}$\alpha$}}
\newcommand{\NII}{\hbox{{\rm [N}\kern 0.1em{\sc ii}{\rm ]}}}
\newcommand{\OIII}{\hbox{{\rm [O}\kern 0.1em{\sc iii}{\rm ]}}}
\newcommand{\NeIII}{\hbox{{\rm [Ne}\kern 0.1em{\sc iii}{\rm ]}}}

\begin{document}
\title{\large \bf Ripples in the OCEANS: Broad Line Variability of Little Red Dots}
\shortauthors{Brooks et al.}
\shorttitle{LRD Variability}


\author[0000-0001-5384-3616]{Madisyn Brooks}
\altaffiliation{NSF Graduate Research Fellow}
\affil{Department of Physics, 196A Auditorium Road, Unit 3046, University of Connecticut, Storrs, CT 06269, USA}

\author[0000-0001-8047-8351]{Kelcey Davis}
\altaffiliation{NSF Graduate Research Fellow}
\affiliation{Department of Physics, 196A Auditorium Road, Unit 3046, University of Connecticut, Storrs, CT 06269, USA}
\affil{Los Alamos National Laboratory, Los Alamos, NM 87545, USA}

\author[0000-0002-1410-0470]{Jonathan R.\ Trump}
\affil{Department of Physics, 196A Auditorium Road, Unit 3046, University of Connecticut, Storrs, CT 06269, USA}

\author[0000-0002-6386-7299]{Raymond C.\ Simons}
\affiliation{Department of Engineering and Physics, Providence College, 1 Cunningham Sq, Providence, RI 02918 USA}

\author[0000-0000-0000-0001]{Erini Lambrides}
\affiliation{Astrophysics Science Division, NASA Goddard Space Flight Center, 8800 Greenbelt Rd, Greenbelt, MD 20771, USA}
\affiliation{Department of Astronomy, University of Maryland, College Park, MD 20742, USA}
\affiliation{Center for Research and Exploration in Space Science and Technology, NASA/GSFC, Greenbelt, MD 20771 USA}


\author[0000-0002-7959-8783]{Pablo Arrabal Haro}
\affiliation{Center for Space Sciences and Technology, UMBC, 5523 Research Park Dr, Baltimore, MD 21228 USA }
\affiliation{Astrophysics Science Division, NASA Goddard Space Flight Center, 8800 Greenbelt Rd, Greenbelt, MD 20771, USA}

\author[0000-0001-8534-7502]{Bren E. Backhaus}
\affil{Department of Physics and Astronomy, University of Kansas, Lawrence, KS 66045, USA}

\author[0000-0001-7151-009X]{Nikko J. Cleri}
\affiliation{Department of Astronomy and Astrophysics, The Pennsylvania State University, University Park, PA 16802, USA}
\affiliation{Institute for Computational and Data Sciences, The Pennsylvania State University, University Park, PA 16802, USA}
\affiliation{Institute for Gravitation and the Cosmos, The Pennsylvania State University, University Park, PA 16802, USA}

\author[0000-0001-8519-1130]{Steven L. Finkelstein}
\affiliation{Department of Astronomy, The University of Texas at Austin, Austin, TX, USA}
\affiliation{Cosmic Frontier Center, The University of Texas at Austin, Austin, TX, USA}

\author[0000-0002-7831-8751]{Mauro Giavalisco}
\affiliation{University of Massachusetts Amherst, 710 North Pleasant Street, Amherst, MA 01003-9305, USA}

\author[0000-0001-9440-8872]{Norman A. Grogin}
\affiliation{Space Telescope Science Institute, 3700 San Martin Drive, Baltimore, MD 21218, USA}

\author[0000-0002-3301-3321]{Michaela Hirschmann}
\affiliation{Institute of Physics, Laboratory of Galaxy Evolution, Ecole Polytechnique Fédérale de Lausanne (EPFL), Observatoire de Sauverny, 1290 Versoix, Switzerland}

\author[0000-0002-8360-3880]{Dale D. Kocevski}
\affiliation{Department of Physics and Astronomy, Colby College, Waterville, ME 04901, USA}

\author[0000-0002-6610-2048]{Anton M. Koekemoer}
\affiliation{Space Telescope Science Institute, 3700 San Martin Drive,
Baltimore, MD 21218, USA}

\author[0000-0003-2366-8858]{Rebecca L. Larson}
\altaffiliation{Giacconi Postdoctoral Fellow}
\affiliation{Space Telescope Science Institute, 3700 San Martin Drive, Baltimore, MD 21218, USA}

\author[0000-0003-1581-7825]{Ray A. Lucas}
\affiliation{Space Telescope Science Institute, 3700 San Martin Drive, Baltimore, MD 21218, USA}

\author[0000-0003-3903-6935]{Stephen M.~Wilkins} %
\affiliation{Astronomy Centre, University of Sussex, Falmer, Brighton BN1 9QH, UK}
\affiliation{Institute of Space Sciences and Astronomy, University of Malta, Msida MSD 2080, Malta}

\author[0000-0003-3735-1931]{Stijn Wuyts}
\affiliation{Department of Physics, University of Bath, Claverton Down, Bath BA2 7AY, UK}

\author[0000-0002-7051-1100]{Jorge A. Zavala}
\affiliation{University of Massachusetts Amherst, 710 North Pleasant Street, Amherst, MA 01003-9305, USA}

\begin{abstract}
Little Red Dots (LRDs) are a unique class of compact, red sources discovered in the JWST extragalactic deep fields. Determining if they are indeed powered by accreting supermassive black holes (SMBHs) is one of the main drivers of the intense study of these objects. Evidence for variability in these objects provides a direct test for the active galactic nucleus (AGN) nature of their central engine. In this study, we present a variability analysis of 6 LRDs observed by the $R \sim 2700$ OCEANS survey and leverage archival $R \sim 1000$ spectroscopic data from the CEERS and RUBIES surveys. We report marginal detections of \Ha\ broad-line (BL) variability in the LRDs OCEANS-100424/RUBIES-42232 (27\% variability at 2.1$\sigma$ significance) and OCEANS-35829/RUBIES-49140 (GlimmIr/Irony; 50\% variability at 1.5$\sigma$ significance). The other 4 LRDs in our sample do not show evidence for BL variability, with a 1$\sigma$ upper limit of $4.8 \% - 30\%$ variability between their epochs of observations. We also find no evidence ($<1\sigma$) for continuum variability in our LRD sample. We compare our results to a sample of SDSS-RM quasars to determine the probability of our broad \Ha\ variability detections. We find that the probability of reproducing 2 variable and 4 nonvariable quasars is $4.71\%$, corresponding to  $\sim 2 \sigma$ departure from typical quasar variability. The detection of BL \Ha\ variability in 2 LRDs provides some evidence for the AGN nature of these objects as opposed to pure scattering models.
\end{abstract}

\section{Introduction}\label{Introduction}

The discovery of the unusual ``little red dot" (LRD: \citealt{Mathee2023, Kocevski2024}) population in the JWST extra-galactic deep fields has led to a flurry of literature investigating the physical mechanisms that drive their unique properties (see \citealt{Inayoshi2025_reviewpaper} for a comprehensive LRD review). These objects are typically described by a compact morphology and a characteristic ``V-shaped" spectral energy distribution (SED) defined by a steep red continuum in the rest-frame optical and elevated blue colors in the UV \citep[e.g.,][]{Mathee2023,Kocevski2024,Kokorev2024, Greene2023, Barro2024, Barro2025, Perez-Gonzalez2026, deGraaff2025b}. The presence of broad (FWHMs $2000-4000 \rm{~km~s^{-1}}$) permitted lines in a substantial fraction ($\sim 80\%$) of LRDs \cite[e.g.,][]{Labbe2023, Kocevski2023,Kocevski2024, Barro2024, Greene2023,Akins2025, Taylor2024, Taylor2025, Lambrides_iron_2025,Lambrides2026, Hviding2025, Wang2025} has been interpreted as a signature of an active galactic nucleus (AGN). LRDs are also found to have strong Balmer breaks \citep[e.g.,][]{Setton2025, Naidu2025, deGraaff2025} that are unusual for AGN and the first studies that reported these strong breaks associated them with massive stellar populations that conflict with $\Lambda \rm{CDM}$ \citep{Labbe2023, Boylan-Kolchin2023}. Additionally, they are usually X-ray faint  \citep{Kocevski2023, Yue2024, Ananna2024, Lambrides2024, Geris2026_xray}, but not always \citep{Hviding2026,Fu2026}, and lack evidence for hot or cold dust emission \citep[e.g.,][]{Williams2024, Perez-Gonzalez2024,Setton2025, Akins2025, Ronayne2026}.

Interestingly, the mechanisms that broaden the hydrogen and helium emission lines observed in LRDs have been highly debated. The Balmer line profiles of some of these sources have been shown to be well described by a combination of a Gaussian core with exponential wings \citep[e.g.,][]{Chang2026,Rusakov2026, Brazzini2026, Kokorev2026, Matthee2026, Davis2026}. This line profile shape can arise from electron (Thompson) scattering in compact, optically-thick ionized cocoons \citep{Laor2006,Rusakov2026}. If the broadening of the permitted emission lines is indeed caused by scattering, the BH masses inferred from single-epoch estimators could be overestimated by 2 orders of magnitude \citep{Brazzini2026}. \cite{Madau2026} has recently shown that the exponential line profile can be modeled with a viralized, radially-stratified BL region (BLR) and \cite{Scholtz2026} analyzed a sample of 32 AGN, including LRDs, and found that exponential profiles are not generally preferred over multi-component Gaussian or Lorentzian profiles. Previous works have shown that the use of non-Gaussian line profiles is common in AGN \citep[e.g.,][]{Kollatschny2013, Kollatschny2018, Scholtz2021, Santos2025}. NGC4395 \citep{Laor2006}, an example of an AGN with an exponential line profile shape, has a reverberation mapping mass measurement \citep{Peterson2005} and a direct dynamical mass measurement \citep{denBrok2015} that are consistent with the virial estimate \citep{Lira1999}. The stratified BLR model can also explain the exponential line profile shapes seen in LRDs without invoking new exotic scattering models \citep{Scholtz2026}.

The physical picture that fully explains the LRD population remains unclear. One interpretation, called the ``Black Hole Star (BH*)", posits LRDs as actively accreting BHs enshrouded in a dense partially ionized gas cocoon \citep[e.g.][]{Inayoshi2025, Naidu2025, Taylor2025, deGraaff2025b, deGraaff2024c, Torralba2026}. Invoking a dense gas shell around a BH can help explain some of the observables that characterize the LRD population \citep{deGraaff2025b, Barro2025, Perez-Gonzalez2026}. \texttt{Cloudy} photoionization models \citep{Ferland2017} of BH*s can reproduce the strong Balmer breaks and high equivalent width (EW) Balmer lines of LRDs \citep[e.g.,][]{Taylor2025, deGraaff2025}. The dense gas cocoon can also explain the narrow absorption lines seemingly superimposed on the broad Balmer emission \citep[e.g,][]{Mathee2023,Matthee2026, Juodzbalis2024, Taylor2024, Lambrides_iron_2025, Lambrides2026, Davis2026} and their large broad-line (BL) Balmer decrements \citep{Brooks2025, deGraaff2025b, Torralba2026}. Recent studies have also suggested a diversity of LRD subtypes, instead of one unified population \citep[e.g.,][]{deGraaff2025b, Perez-Gonzalez2026, Barro2025, Billand2026}. This could complicate the LRD picture further, as different physical mechanisms could potentially describe the different LRD subtypes.  

Determining if LRDs are indeed powered by accreting super-massive BHs (SMBHs), perhaps enshrouded by dense-gas shells, is critical for our comprehensive understanding of BH-galaxy co-evolution in the early Universe. Scaling relations applied to high-$z$ AGN and LRDs are derived from the local Universe \citep[e.g.,][]{Greene2005, ReinesVolonteri2015, DallaBonta2025}, which rely on reverberation mapping (RM) \citep{Blandford1982} measurements of AGN variability.  This AGN variability is commonly seen on month-to-year timescales \citep[e.g.,][]{Clavel1992,VandenBerk2004,Kelly2009, Macleod2012,Edelson2015}, now even being studied at cosmic dawn \citep{Leung2026}. If LRDs are indeed powered by an AGN, we would expect evidence for variability driven by the time-lag between the AGN continuum and the photo-ionization of the BL region. Probing the variability of LRDs would provide a robust constraint on the physical mechanisms that power them. Detecting variability in LRDs would imply a direct sightline to the accretion disk or BLR of the AGN.

Surprisingly, the majority of LRDs lack clear evidence for variability in both spectroscopic and photometric surveys \citep[e.g.,][]{Zhang2025, Kokubo2025, Burke2025, Tee2025, Liu2026, Stone2026_nircam,Stone2026} in 3D radiation magnetohydrodynamic simulations of AGN disks \citep{Secunda2026}, and in local analogs \citep{Lin2026}. Unique cases of variability have been reported in LRDs, such as in an over-massive multiply-imaged LRD \citep{Furtak2025, Ji2025}, and ``MoM-BH*-1" \citep{Naidu2025}. Both narrow and broad-line variability have also been reported for the GlimmIr, a $z\sim7$ LRD \citep{Lambrides2026}. The infrequent detections of LRD variability do not yet establish if variability is a common property of the LRD population or if it is limited to a small subset of extreme sources. Expanding the sample of LRDs with repeat spectroscopy is necessary for fully understanding the accretion variability in this population. 

In this paper, we present a variability analysis of a subset of LRDs observed by the OCEANS survey. We explore continuum and broad \Ha\ line flux variability for a sample of 6 LRDs using new high-resolution NIRSpec ($R \sim 2700$) data from the OCEANS survey and archival medium-resolution NIRSpec ($R \sim 1000$) data from the CEERS and RUBIES surveys. We report marginal detections of \Ha\ broad-line (BL) variability in the LRDs OCEANS-100424/RUBIES-42232 and OCEANS-35829/RUBIES-49140 (Irony/GlimmIr; \citealt{Deugnio2026_irony, Lambrides_iron_2025, Lambrides2026}), referred to here as the GlimmIr. We also derive upper limits on the \Ha\ BL variability for the sources with no detected \Ha\ BL variability. We perform a comparison of our LRD variability to a sample of SDSS-RM quasars \citep{Shen2024} to determine the likelihood of LRDs to have consistent variability properties to low-redshift accreting supermassive BHs.

The paper is presented as follows. In \S \ref{sec:Spectroscopic Data} we describe our sample selection and observational data set, including the NIRSpec data reduction for the high-resolution OCEANS data. In \S \ref{sec: methods} we describe our multi-epoch flux calibration and emission line fitting methodology. In \S \ref{sec: results} we present our results and discuss our findings. In \S \ref{sec: conclusions} we describe future work. 
For this study, we assume a flat $\mathrm{\Lambda}$CDM cosmology with $\mathrm{H_{0}}$ = 67.4 km s $^{-1}$ Mpc$^{-1}$ and $\Omega_{\mathrm{M}} = 0.315$ \citep{Planck2020}.

\section{Observational Data}\label{sec:Spectroscopic Data}

\begin{figure}
    \centering
    \includegraphics[width=\linewidth]{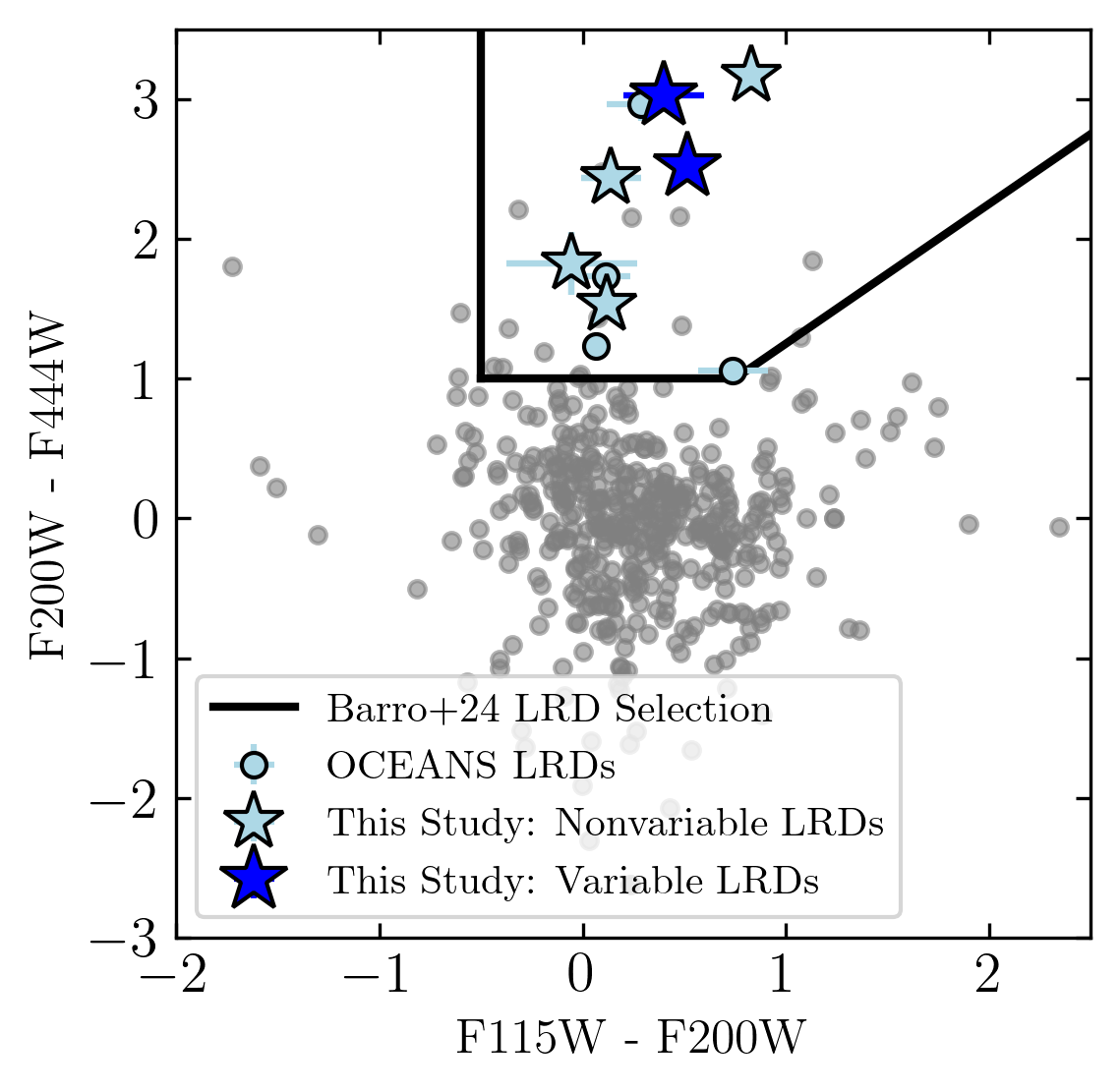}
    \caption{F155W-F200W vs F200W-F444W for the full OCEANS sample. The full OCEANS sample is shown with the gray circles. Targets selected to be LRDs following the prescription described in \cite{Barro2024, Barro2025} are shown with the light blue circles. LRDs that are included in this work are shown with stars. Two LRDs, OCEANS-100424 and the GlimmIr, with marginal evidence for broad \Ha\ variability are shown with the dark blue stars.}
    \label{fig:lrd selection}
\end{figure}

In this study, we utilize JWST NIRSpec/MSA spectroscopy from the Observing Cosmic Evolution Among Nascent Systems (OCEANS) survey (GO-8410, PI: Raymond Simons), the Cosmic Evolution Early Release Science Survey (CEERS) (ERS-1345, PI: Steven Finkelstein; \citealt{Finkelstein2025}), and the Red Unknowns: Bright Infrared Extragalactic Survey (RUBIES) (GO-4233, PIs: Anna de Graaff, Gabriel Brammer; \citealt{deGraaff_rubies_survey}) to study the BL variability of the LRD population. We additionally use NIRCam photometry from CEERS for our LRD selection, where we follow the color-color selection prescription described in \citep{Barro2024,Barro2025}. Figure \ref{fig:lrd selection} shows the sources analyzed in this paper and their position in color-color space. We also include the LRD selection criteria described in \cite{Barro2024}. Table \ref{tab:sources} lists the OCEANS NIRSpec ID, previous survey IDs, and the RAs and DECs of the sources in this study.

We select sources for our variability analysis by their \Ha\ and \OIII\ spectral coverage in previous spectroscopic surveys. We include sources with previous \OIII\ or \NeIII\ coverage to flux-calibrate the multiple epochs of spectroscopy, described in \S \ref{sec: flux calibration}. Since the RUBIES survey only observed with the G395M/F290LP grating/filter pair, this limits the viable sources to the redshift range $4.90<z<6.77$. In total, we study 6 LRDs re-observed in the OCEANS survey. A brief description of the observational programs used in this study follows.

\begin{deluxetable*}{cccccc}[]

\tablecaption{OCEANS LRD Sources \label{tab:sources}}

\tablehead{
\colhead{OCEANS ID} &
\colhead{Alternate ID} &
\colhead{R.A.} & 
\colhead{Dec} &
\colhead{$z$} & 
\colhead{Time between Observations} \\
\colhead{} &
\colhead{} & 
\colhead{} & 
\colhead{} & 
\colhead{} &
\colhead{[rest-frame days]}
}
\startdata 
100424* & RUBIES-42232 & 214.886792 & 52.855381 & 4.953 & 121\\
169045 & RUBIES-50052/CEERS-2782 & 214.823454 & 52.830277 &  5.239 & 72, 187\\
20504 & RUBIES-42046 & 214.795368 & 52.788847 & 5.276  & 114\\
33842 & RUBIES-60935 & 214.923373 & 52.925593 & 5.287 & 116\\
161695 & CEERS-672 & 214.889677 & 52.832977 & 5.666 & 175\\
35829 (GlimmIr)* & RUBIES-49140 & 214.892249 & 52.877403 & 6.684 & 101
\enddata
\centering
\tablecomments{Sources denoted with a * show marginal evidence for broad \Ha\ variability.}
\end{deluxetable*}

\subsection{OCEANS Spectroscopy}\label{sec: oceans spectra}
The OCEANS survey observed 6 NIRSpec MSA pointings in the CEERS (EGS) field. The pointings were observed with the G235H/F170LP and G395H/F290LP grating/filter pairs, providing $R \sim 2700$ spectral coverage from $1.7 - 5.1 ~\mu m$. Each NIRSpec pointing was observed for $\sim$4.1~hr per grating. Sources observed in multiple NIRSpec pointings could reach up to $\sim$12~hr of total integration time.

The OCEANS survey is designed to study both the formation of galactic disks in the early Universe and the kinematics/compositions of galactic outflows across cosmic time (Simons et al.\ in prep). OCEANS also targets multiple LRD and BLAGN sources. The spectroscopic targets for the OCEANS survey were selected from NIRCam imaging in the CEERS survey \citep{Finkelstein2025} and NIRSpec low/medium resolution spectroscopy from CEERS, CANDELS-Area Prism Epoch of Reionization Survey (CAPERS) (GO-6368; PI: M. Dickinson), and The High-(Redshift+Ionization) Line Search (THRILS) (GO-5507; PIs: T. Hutchison and R. Larson; \citealt{Hutchison2025}), and RUBIES (GO-4233; PIs:
A. de Graff and G. Brammer; \citealt{deGraaff_rubies_survey}).

LRD and AGN sources targeted through the OCEANS survey were prioritized through the following scheme: (1) spectroscopic confirmation of broad Balmer emission \citep{Taylor2024}, photometric LRD selection \citep{Kocevski2024, Taylor2024}, and Balmer absorption \citep{Taylor2024} or (2) photometric LRD selection or targets with previous spectroscopic epochs \citep{Kocevski2023, Kocevski2024, Taylor2024}.

Three LRDs, OCEANS-20504, OCEANS-161695, and the GlimmIr, were observed with multiple NIRSpec configurations. The GlimmIr only had \Ha\ observed on the NIRSpec detector (and not in a chip gap) in P6, so we use that observation for our analysis of its \Ha\ line profile. P6 also coincidentally has a chip-gap in its  \OIII $\lambda\lambda4959,5007$ and $5100 \rm \AA$ region, which we use for our \OIII\ flux and continuum luminosity measurements, described in \S \ref{sec: methods}. For these measurements of this source, we co-add the 2 other OCEANS observations (P2 and P3) and use that as our 1D spectra for the \OIII\ region. For analysis of OCEANS-20504 and OCEANS-161695, we co-add the 1D spectra to improve the signal-to-noise. To co-add the multiple epochs of observation, we re-sample the spectra to the same wavelength grid and then take the inverse-variance weighted average at each wavelength-pixel position. The error on the co-add is estimated as $\sqrt{\frac{Var(x)_{wtd}}{n}}$, where $Var(x)_{wtd}$ is the weighted variance and n is the number of pixels. Bad pixels and chip-gaps are also masked out from the final co-added spectra. 

\subsubsection{OCEANS NIRSpec Data Reduction}
The NIRSpec MOS data from the OCEANS survey was processed with the STScI JWST Calibration Pipeline v1.20.2 (CRDS context jwst\_1464.pmap; \citealt{Bushouse2025}). The NIRSpec MOS pipeline notebook \citep{jwst_pipeline_notebooks} was adapted for cosmic ray rejection and bad-pixel self calibration. In Stage 1 (\texttt{calwebb\_detector1}), we enhance the snowball correction factor to an expansion factor of 3 and a minimum flagged area of 15 pixels. In Stage 2 of the pipeline (\texttt{calwebb\_spec2}), we enable the bad pixel self-calibration (\texttt{badpix\_selfcal}; $f_\mathrm{flag} = 0.005$). We then apply the standard nodded background subtraction. Final spectra produced in Stage 3 (\texttt{calwebb\_spec3}) were produced by co-adding the two-nod positions for each source. The 1D spectra were then extracted using a 5-pixel wide box-car centered on the source.

Previous studies \citep[e.g,][]{Maseda2023} have shown that the uncertainties computed by the NIRSpec JWST pipeline are frequently underestimated. To mitigate this effect in our data, we calculate empirical correction factors for our extracted 1D spectra. We first remove the overall continuum shape by modeling it with a 3rd-order polynomial and subtracting it from the 1D spectra. We then calculate the median absolute deviation (MAD) of the continuum-subtracted spectra. The MAD is multiplied by 1.4826 to give the ``empirical scatter" of our data. The correction factor that is applied to the NIRSpec uncertainties is found by taking the ratio of the empirical scatter with the median of the pipeline reported errors. We find that the NIRSpec pipeline systematically underestimates the flux uncertainties of our data by $\sim 35-65\%$. 

\subsection{Archival Spectroscopy}

\begin{figure*}[t]
 \centering
    \includegraphics[width=\linewidth]{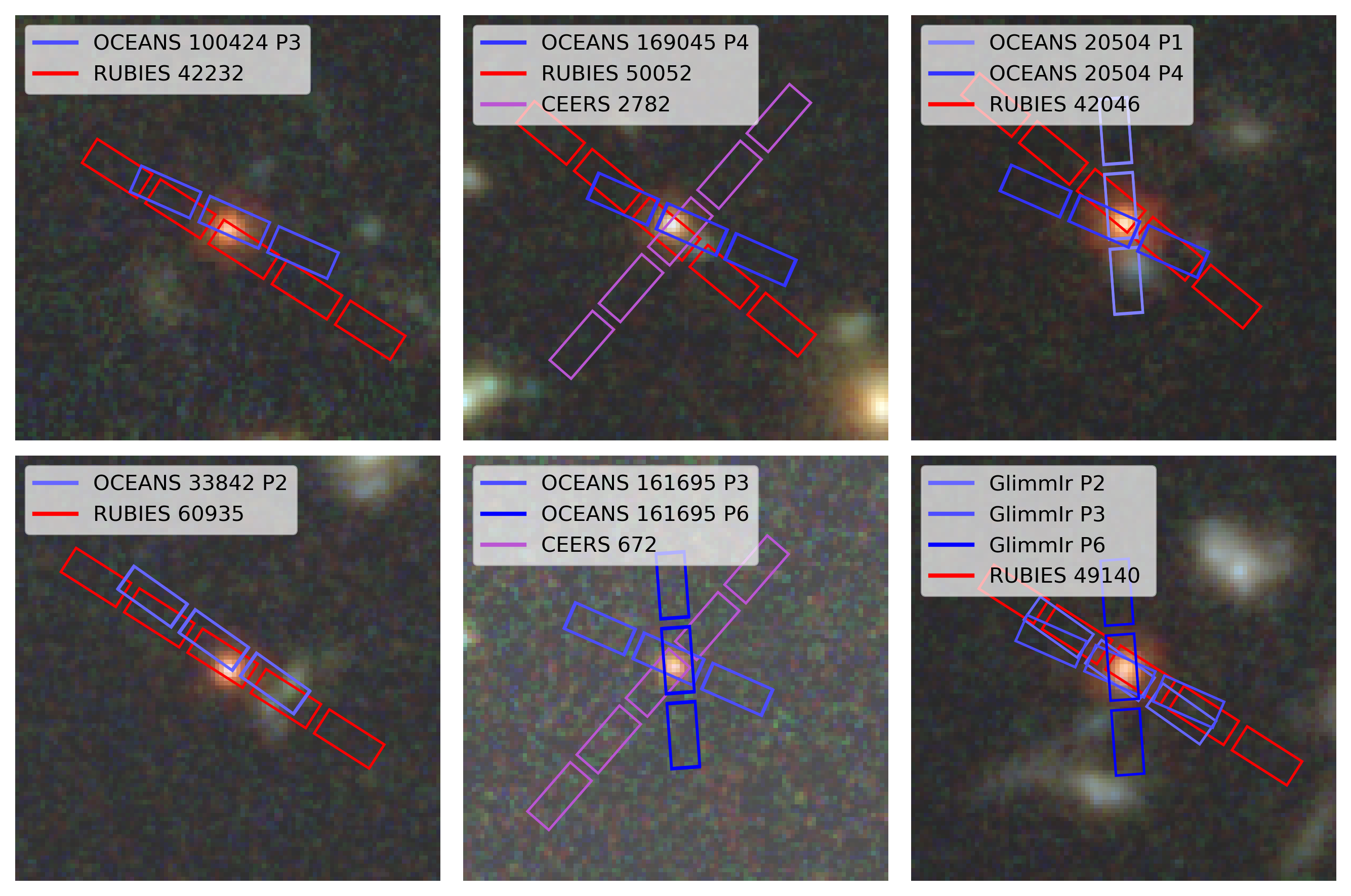}
    \caption{Color image cutouts and NIRSpec MSA slit configurations for our sample of LRDs. OCEANS configurations are shown in blue, RUBIES in red, and CEERS in purple. All images are $3^{\prime \prime} \times 3^{\prime \prime}$ in size.}
    \label{fig:bio plots}
\end{figure*}

We utilize archival NIRSpec medium-resolution ($R\sim 1000$) grating spectroscopy from the CEERS survey \citep{Finkelstein2025} and the RUBIES survey \citep{deGraaff_rubies_survey}. The CEERS survey observed 6 NIRSpec pointings of the Extended Growth Strip (EGS); the pointings were observed with the G140M/F100LP, G235M/F170LP, and G395M/F290LP grating/filter pairs, and for 0.86~hr per grating. In this study, we use CEERS data only from the G395M/F290LP grating/filter pair. The RUBIES survey also observed 6 NIRSpec pointings in the EGS field with the G395M/F290LP grating/filter pair and with an exposure time of 0.80~hr. We refer to \citet{Finkelstein2025} and \cite{deGraaff_rubies_survey} for a full description of the CEERS and RUBIES surveys, respectively. The publicly available archival spectroscopy used in this study was retrieved through the DAWN JWST Archive (DJA) \citep{DJA}. We refer to \cite{deGraaff_rubies_survey} and \cite{Heintz2024,Heintz2025} for a full description of the NIRSpec data reduction. The NIRSpec MSA configurations for the different spectroscopic observations are shown in Figure \ref{fig:bio plots}.

OCEANS-100424 was observed in two separate RUBIES pointings, P62 and P63. We elect to drop the P63 of OCEANS-100424 from this study due to artifacts present in the 2D spectra that are particularly apparent in the \Ha\ region. We have measured the spatial extent of the \Ha\ region in the OCEANS, RUBIES P62, and RUBIES P63 spectra, and find a broader spatial size in the cross-dispersion direction for the narrow \Ha\ emission component in the RUBIES P63 2D spectra while the OCEANS and RUBIES P62 spectra have identical spatial profiles. We further describe our analysis and issues with the RUBIES P63 spectra in the Appendix. The specific observations analyzed in this paper can be accessed via \dataset[doi: 10.17909/tj0h-cj44]{https://doi.org/10.17909/tj0h-cj44}.

\section{Methodology}\label{sec: methods}

\subsection{Flux Calibration}\label{sec: flux calibration}
In order to compare our multiple epochs of spectroscopy and to account for different slit orientations and aperture losses, we flux calibrate our spectra by the \OIII\ flux ratio between the different epochs. The \OIII $\lambda\lambda4959,5007$ doublet is fit with an MCMC fitting routine that assumes a single narrow Gaussian emission line component for each line. The \OIII\ line fluxes are fixed to a ratio of 1:2.985 \citep{Storey2000}. The BLAGN included in our sample have been previously checked for broad \OIII\ emission which would be indicative of a galactic outflow scenario driving the BL kinematics instead of a BLAGN \citep{Kocevski2023,Taylor2024, Brooks2025, Davis2026}. \cite{Taylor2024} found evidence for a broad \OIII\ component in RUBIES-50052/CEERS-2782 but with a more weaker and narrower velocity width than the broad \Ha\ component. The difference in the broad FWHMs of this source indicates a BLAGN in this system, so we elect to keep it in our sample. No other evidence for galactic outflows in the \OIII\ lines is found in our sample. 

The P6 observation of the GlimmIr suffers from a chip gap in its \OIII\ line region. For this source, we opt to use the $\NeIII \lambda 3869$ emission line for our flux calibration. We note that the spatial resolution of the NIRSpec high-resolution grating is wavelength dependent, which could introduce a systematic offset between an \OIII -based and a \NeIII -based flux calibration. We find that the \NeIII\ flux ratio between P6 and the RUBIES observation is consistent with the P3 and RUBIES \OIII\ flux ratio, indicating that the wavelength dependence of the high-resolution grating does not introduce a substantial systematic offset in our calibration. We expect the P2 and P3 \OIII\ fluxes of the GlimmIr to be consistent due to their similar slit placements. 

\begin{figure*}
     \centering
    \includegraphics[width=\linewidth]{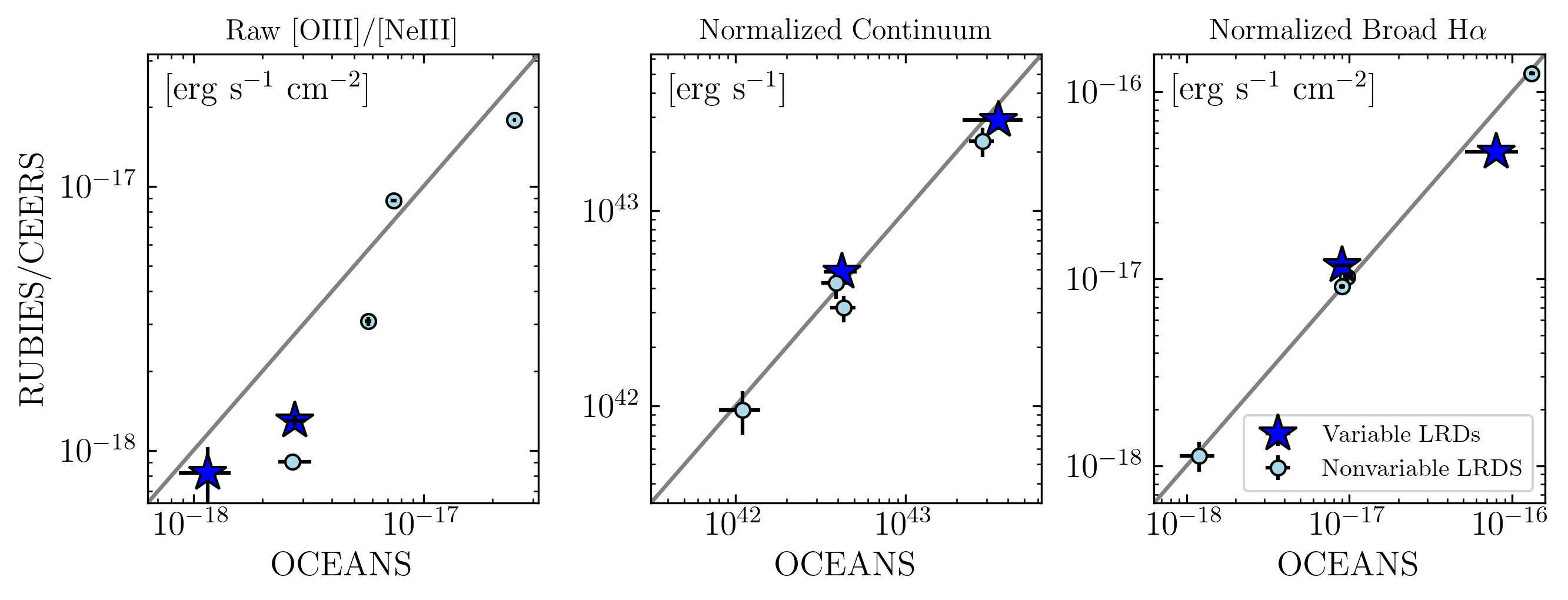}
    \caption{OCEANS vs RUBIES/CEERS flux and luminositiy  measurements. The left panel shows the raw \NeIII\ flux for the GlimmIr and the raw \OIII\ flux for the other sources in the sample, the middle panel shows the normalized $5100 \rm \AA$ continuum luminosity, and the right panel shows the normalized broad \Ha\ flux. We use the \NeIII\ or \OIII\ flux ratio to calibrate the RUBIES/CEERS spectra, as described in \S \ref{sec: flux calibration}. Our sources that show evidence for broad \Ha\ variability, OCEANS-100424 and the GlimmIr, are denoted with the dark blue stars. The nonvariable LRDs included in this study are denoted with the light blue circles.}
    \label{fig: flux measurements}
\end{figure*}

Once the narrow \OIII\ (or narrow \NeIII) emission line flux is measured for both the OCEANS epoch and the previous spectroscopic epoch(s), we normalize the previous epoch(s) by the \OIII\ or \NeIII\ flux ratio with the following:
\begin{equation}
    F_{\lambda,\mathrm{normalized}}
    =
    F_\lambda \times \frac{F_{narrow} ~{\rm OCEANS}}{F_{narrow}{ ~\rm RUBIES/CEERS}}
\end{equation}
We opt to use the \OIII\ or \NeIII\ line flux as a flux calibration standard, as it has been shown to be constant over the rest-frame $\sim$1~yr timescales used in this study and only observed to vary on rest-frame decades \citep{Foltz1981,Peterson1982,Peterson2013, Fries2023}. We note that the constant \OIII\ fluxes from these sources are in part due to a well developed narrow line region with a large extent, which might not be applicable in LRDs \citep{Ishikawa2026}. The \OIII\ or \NeIII\ emission line measurements from the raw spectra are shown in Figure \ref{fig: flux measurements} and the normalization ratios are reported in Table \ref{tab: flux normalization}.  We add the normalization ratio error in quadrature to all flux estimations where the normalization is applied.

\begin{deluxetable}{cc}
\label{tab: flux normalization}
\tablehead{
\colhead{OCEANS ID} & 
\colhead{Epoch Flux Ratio (\OIII\ or \NeIII\ )}}
\caption{Flux Calibrations}

\startdata
100424 & $0.47 \pm 0.06$ \\
169045 & $0.72 \pm 0.01$ \\
20504 & $1.19 \pm 0.04$ \\
33842 & $0.54 \pm 0.03$ \\
161695 & $0.34 \pm 0.06$ \\
35829 & $0.71 \pm 0.25$ \\
\enddata
\tablecomments{OCEANS-35829 uses \NeIII\ for its flux calibration, while the rest of the LRD sample uses \OIII\ .}
\end{deluxetable}

\subsection{H$\alpha$ Emission Line Fitting}\label{sec: emission line fitting}

\begin{figure*}
    \centering
    \includegraphics[width=\linewidth]{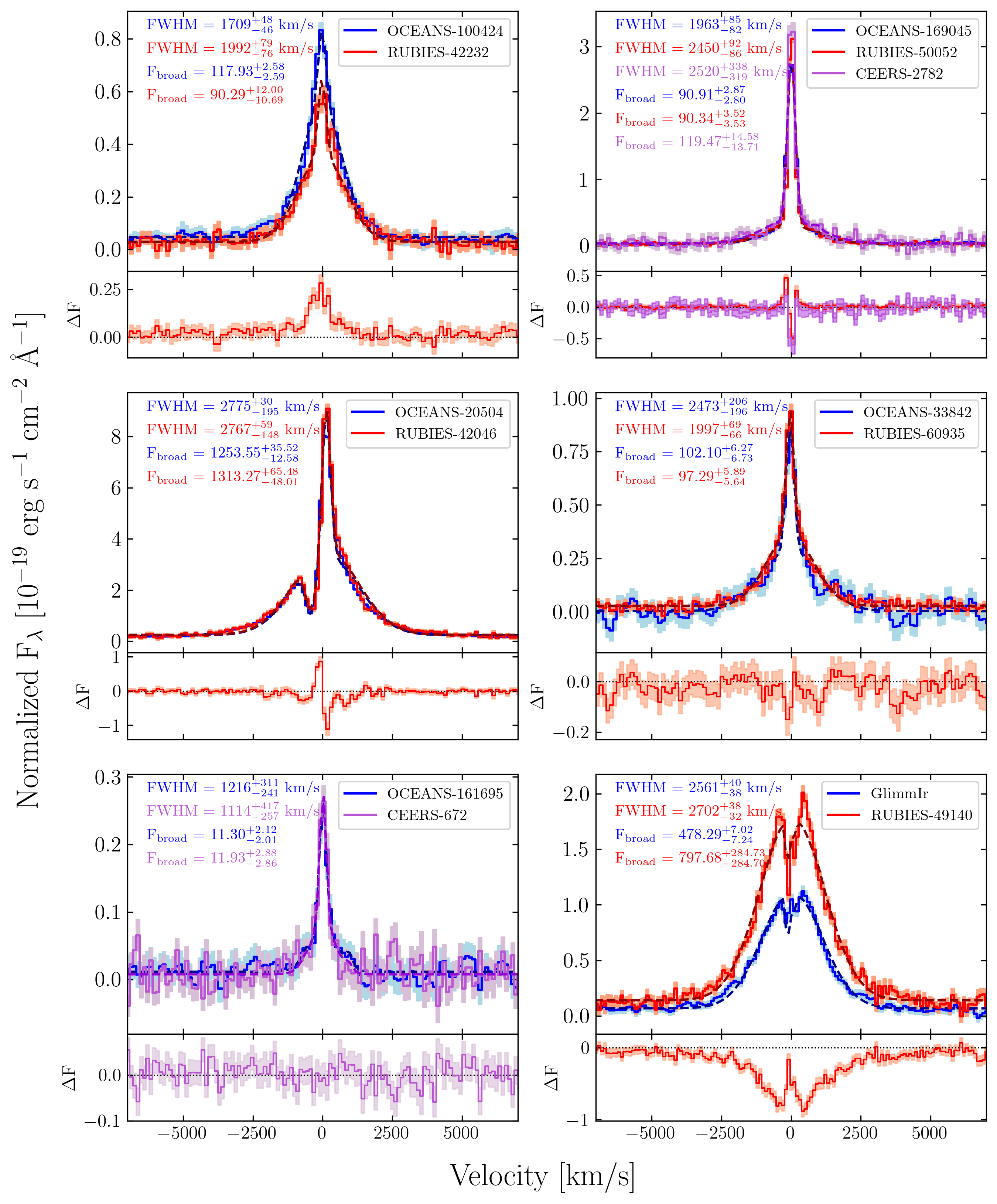}
    \caption{Multi-epoch \Ha\ line profiles for the OCEANS LRD sample. OCEANS observations are shown in blue, CEERS observations are shown in purple, and RUBIES observations are shown in red. The CEERS and RUBIES observations shown are flux calibrated by their \OIII\ emission line flux, described in \S \ref{sec: flux calibration}. The emission line fits, described in \S \ref{sec: emission line fitting}, are shown with the dashed curves. The \Ha\ BL FWHM and BL fluxes for the different epochs are shown in the upper left corner. BL fluxes are reported in units of $10^{-19}~ \rm{erg ~s^{-1} ~cm^{-2}}$.}
    \label{fig: spectra}
\end{figure*}

After the spectra are flux-calibrated, we simultaneously fit the $\Ha + \NII$ complex for all observation epochs using the Markov Chain Monte Carlo (MCMC) routine from the Python \texttt{emcee} package \citep{emcee}. We implement a dual-component narrow+broad Gaussian model for the \Ha\ emission line, building off the fitting technique described in \cite{Brooks2025, Brooks2026}. We first use the $R \sim 2700$ OCEANS spectra to determine $\sigma_{nar}$ from the $\lambda\lambda4959,5007$ doublet emission line fit. The multiple spectroscopic epochs are then forced to share the same total narrow \Ha\ line flux; the $R \sim 1000$ spectra use the $\sigma_{nar}$ determined from the high-resolution \OIII\ line fit with the $R \sim 1000$ resolution added in quadrature to broaden the intrinsic $\sigma_{nar}$. This allows the amplitudes of the individual narrow \Ha\ components to differ while conserving the same total narrow \Ha\ line flux. We model the \NII$\lambda\lambda$6550,6585 doublet with a fixed amplitude ratio of 1:2.94 \citep{Osterbrock2006} and fix the \NII\ line width and line center to the narrow \Ha\ emission component. We then implement the following priors in our \Ha\ fitting routine: positive line fluxes for both the narrow and broad emission line component and $\rm{FWHM_{broad} > 700 ~km~s^{-1}}$. The BL center is fixed to the narrow line center, which is allowed to vary within $250 \rm{~km~s^{-1}}$ of the line center determined by the \OIII\ redshift. We run \texttt{emcee} using 32 walkers for 20,000 steps, and implement``burn-in" of 10,000 steps that are discarded from our final analysis. For OCEANS-169045, which has both a previous CEERS and RUBIES observation, we follow the same procedure as above but with an additional 4 fit parameters to account for the additional spectroscopic epoch. The \Ha\ emission line profiles and best-fit models for our sample are shown in Fig \ref{fig: spectra}.

\begin{deluxetable*}{llcccccc}
\tablewidth{0pt}
\tablecaption{Multi-epoch Broad H$\alpha$ Emission Line and Continuum Properties \label{tab:measurements}}
\tablehead{
\colhead{} & \colhead{} & \colhead{100424*} & \colhead{169045} & \colhead{20504} & \colhead{33842} & \colhead{161695} & \colhead{35829 (GlimmIr)*}
}
\startdata
\multirow{3}{*}{$F_{\rm broad}$}
 & OCEANS & $117.9^{+2.6}_{-2.6}$ & 
$91.03^{+2.9}_{-3}$ & 
$1254^{+36}_{-13}$ & 
$102.1^{+6.3}_{-6.7}$ & 
$11.3^{+2.1}_{-2}$ & 
$478.3^{+7}_{-7.2}$ \\
 & CEERS & $-$ &
$119.5^{+15}_{-14}$ & 
$-$ &
$-$ &
$11.93^{+2.9}_{-2.9}$ & 
$-$   \\
 & RUBIES & $90.29^{+12}_{-11}$ & 
$90.34^{+3.5}_{-3.5}$ & 
$1313^{+65}_{-48}$ & 
$97.29^{+5.9}_{-5.6}$ & 
$-$ &
$797.7^{+280}_{-280}$    \\
\hline
\multirow{3}{*}{FWHM}
 & OCEANS & $1709^{+48}_{-46}$ & 
$1965^{+86}_{-84}$ & 
$2775^{+30}_{-190}$ & 
$2473^{+210}_{-202}$ & 
$1216^{+310}_{-240}$ & 
$2561^{+40}_{-38}$ 
 \\
 & CEERS & $-$ &
$2513^{+350}_{-320}$ & 
$-$ &
$-$ &
$1114^{+420}_{-260}$ & 
$-$  \\
 & RUBIES & $1992^{+79}_{-76}$ & 
$2452^{+92}_{-88}$ & 
$2767^{+59}_{-150}$ & 
$1997^{+69}_{-66}$ & 
$-$ &
$2702^{+38}_{-32}$ \\
\hline
\multirow{3}{*}{$L_{5100}$}
 & OCEANS & $0.48 \pm 0.08$ & 
$0.42 \pm 0.07$ & 
$2.27 \pm 0.39$ & 
$0.32 \pm 0.05$ & 
$0.09 \pm 0.02$ & 
$2.91 \pm 0.32$ \\
 & CEERS & $-$ &
$0.45 \pm 0.10$ & 
$-$ &
$-$ &
$0.11 \pm 0.03$ & 
$-$   \\
 & RUBIES & $0.42 \pm 0.09$ & 
$0.39 \pm 0.07$ & 
$2.83 \pm 0.47$ & 
$0.43 \pm 0.07$ & 
$-$ &
$3.52 \pm 1.36$  \\
\enddata
\tablecomments{$F_{\rm broad}$ is given in units of $10^{-19}\,{\rm erg\,s^{-1}\,cm^{-2}}$, FWHM in km s$^{-1}$, and $L_{5100}$ in units of $10^{43}\,{\rm erg\,s^{-1}}$. Sources denoted with a * show marginal evidence for broad \Ha\ variability.}
\end{deluxetable*}

Two of our OCEANS LRDs have Balmer absorption superimposed on their broad Balmer emission lines \citep{Davis2026}. For these sources, OCEANS-20504 and the GlimmIr, we adapt the above fitting procedure to include an absorption line component, which is modeled with a single Gaussian ($f_{abs}$, $\sigma_{abs}$, and $\lambda_{abs}$). The different epochs are then forced to have the same absorption line properties. 

\citet{Davis2026} reports evidence for a weak redshifted \Ha\ absorption component ($\rm EW = 1.1 \AA$) in OCEANS-100424. We fit the RUBIES $R \sim 1000$ spectra for this source with a dual narrow+broad Gaussian with an additional Gaussian to model the absorption component. In the RUBIES spectra, we do not find evidence that an additional absorption component improves the emission line fit. The absorption line parameters are also not well constrained. For the variability study of OCEANS-100424, we elect to model its \Ha\ emission line region with only the dual narrow+Gaussian model and without any absorption component.

To further test our fitting methodology, we degrade the OCEANS $R \sim 2700$ spectra to $R \sim 1000$ by convolving the spectra with the difference in line spread functions of the two spectral resolutions. We then re-sample the OCEANS spectra onto the same wavelength grid as the $R \sim 1000$ data. We then follow the same fitting procedure described above, but now the multi-epochs share the same $\sigma_{nar}$, which is determined from the $R \sim 2700$ \OIII\ fit and then broadened to match the $R \sim 1000$ spectral resolution. We find that this alternate fitting procedure is statistically consistent with our first methodology. The emission line measurements we report in this study use the first fitting procedure described, which leverages the $R \sim 2700$ OCEANS spectra.

We do not check for \Hb\ variability in our sample due to the low SNR of the broad \Hb\ emission line, especially in the RUBIES $R \sim 1000$ spectroscopy. Broad \Hb\ is only detected in 3/6 of our OCEANS $R \sim 2700$ sample \citep{Davis2026}. See \citet{Brooks2025} for an analysis of the broad Balmer decrements of BLAGN which contributes to the lack of high SNR measurements of broad \Hb\ . 

\subsection{Continuum Luminosity Measurements}\label{sec: continuum luminosity}

We measure the continuum luminosity in the $5050-5200 \rm \AA$ region for the OCEANS spectra and the CEERS/RUBIES \OIII\ flux-calibrated spectra to check for continuum variability of our LRD sample. We take the median of the flux in this region and use it as our $5100 \rm \AA$ flux measurement and we estimate the error on the median as $1.253 \times \sigma_{\Bar{x}}$.

For the GlimmIr, \Ha\ was observed on the NIRSpec detector in P6, and not in its other NIRSpec configurations, P2 and P3. P6 also suffers from a chip gap in its spectra in the $5100 \rm \AA$ angstrom region. To measure the continuum of this source, we use the co-added spectra of P2 and P3, described in \S \ref{sec: oceans spectra}. The individual OCEANS pointings of the GlimmIr are discussed in more detail in Lambrides+2026b.



\section{Results and Discussion}\label{sec: results}
\begin{figure*}
    \centering
    \includegraphics[width=.8\linewidth]{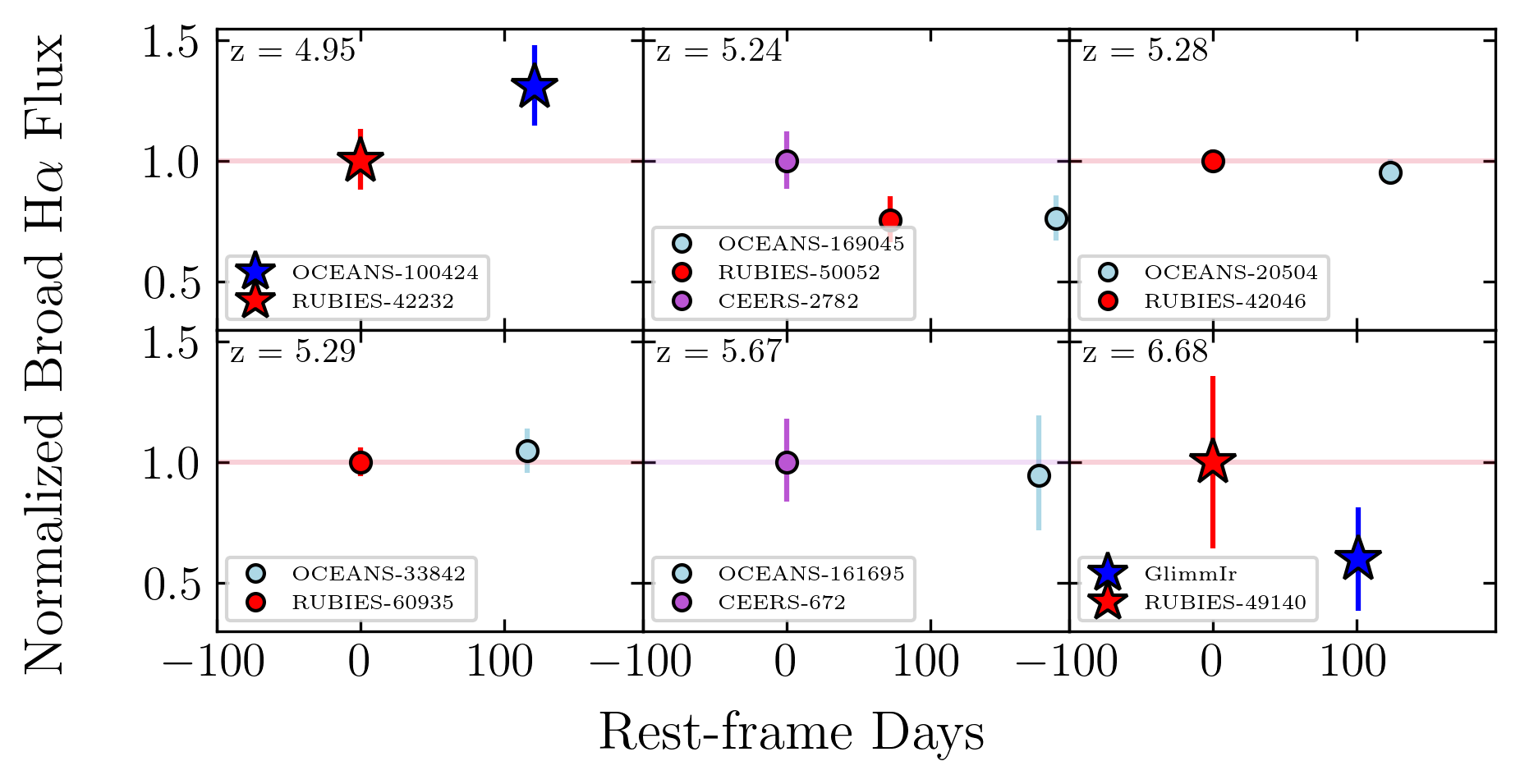}
    \caption{Normalized broad \Ha\ flux versus rest-frame time between spectroscopic observations for the 6 OCEANS LRDs included in this study. The OCEANS observations are shown in blue, the RUBIES observations are shown in red, and the CEERS observations are shown in purple. The two sources with evidence for broad \Ha\ emission line variability, OCEANS-100424/RUBIES-42232 and GlimmIr/RUBIES-49140, are denoted with stars. The other 4 LRDs in this study are denoted with circles.}
    \label{fig: rest frame days}
\end{figure*}

In this study, we search for evidence of \Ha\ BL and continuum variability in a sample of 6 LRDs from the OCEANS spectroscopic survey. These sources cover a redshift range of $4.95 < z < 6.68$ and have rest-frame time differences in their observation epochs of 72-187 days. We measure a flux ratio of $\rm \frac{F_{OCEANS}}{F_{RUBIES}} = 1.31^{+0.16}_{-0.18}$ for OCEANS-100424 and $\rm \frac{F_{OCEANS}}{F_{RUBIES}} = 0.60^{+0.21}_{-0.21}$ for the GlimmIr. We measure the variability amplitudes of these two sources to be $27\%$ ($2.1\sigma$) and $50\%$ ($1.5\sigma$), respectively.

The low significance of both the OCEANS-100424 and the GlimmIr variability detection are driven by the large uncertainties on the flux normalization ratio, described in \S \ref{sec: flux calibration}. The GlimmIr is especially affected by the normalization uncertainty due to the low SNR of the \NeIII\ measurements. The \OIII\ normalization ratio for the P3 OCEANS and RUBIES epoch has 25x greater sensitivity than the P6 \NeIII\ ratio used here. The \NeIII\ normalization ratio is adopted in this work due to the \NeIII\ coverage in the P6 observation of the GlimmIr (the same observation with \Ha\ coverage). 
The normalization uncertainty accounts for the majority of the total flux uncertainty in the GlimmIr's RUBIES epoch measurement, with the broad \Ha\ emission line flux uncertainty contributing negligibly by comparison. This indicates that the low significance ($1.5\sigma$) of the GlimmIr's variability is primarily a consequence of the normalization uncertainty rather than intrinsically weak variability. 

The remaining 4 LRDs included in this study do not show any evidence for broad \Ha\ variability, with variability amplitudes ranging from $0.8 \% - 5.4\%$ ($0.15 - 0.96 \sigma$). We derive $1\sigma$ variability amplitude upper limits as $\frac{\sigma_{\Delta F}}{\bar{F}}$, and find a $1\sigma$ upper limit of $4.8 \% - 30 \%$ for the four nonvariable LRDs.
The broad \Ha\ line fluxes and FWHMs of the 6 LRDs are given in Table \ref{tab:measurements}.

We additionally do not find any evidence for significant ($> 3\sigma$) continuum variability in our sample. Figure \ref{fig: flux measurements} (center panel) shows the continuum luminosity measurements for our sample. We show the broad \Ha\ flux versus rest-frame time for our sample in Figure \ref{fig: rest frame days}. For a more detailed study of the Balmer profiles and variability properties of the GlimmIr, see Lambrides+2026b in prep.



\subsection{What does it mean for LRDs to be variable?}

The physical mechanisms that power LRDs still remain highly uncertain. They display broad Balmer emission lines with FWHMs $> 1000 \rm ~km~s^{-1}$, which is typically associated with a rapidly rotating gas around a central SMBH. They are also extremely compact, with effective radii $< 200 ~ \rm pc$ \citep[e.g.,][]{Baggen2023, Baggen2024, Akins2025}, consistent with the point-like nature of AGN. AGN BL and continuum emission are well known to vary over time-scales of hours to decades \citep[e.g.,][]{VandenBerk2004, Kelly2009, Kozlowski2010, MacLeod2010, Macleod2012}. Observing variability in LRDs would provide a key observational diagnostic for their AGN nature. 

If no variability is observed in the Balmer emission lines, this could suggest a non-AGN explanation for LRDs, like extremely compact, massive galaxies \citep{Baggen2024}, globular clusters \citep{Chisholm2026}, or super-massive stars \citep{Nandal2026}. The lack of observed variability could also suggest that the LRD is a BH enshrouded in a dense gas cocoon \citep[e.g.,][]{Inayoshi2025_bhstar} with scattering dominated emission \citep[e.g.,][]{Rusakov2026, Kokorev2026, Matthee2026}. In this framework, the observed broad emission lines are produced through electron scattering of the photons into the line of sight. Any variability in the BL region will be reprocessed and smoothed by the electron scattering in the dense gas medium. 

Super-Eddington accretion can also explain the lack of variability observed in LRDs \cite[e.g.,][]{Lambrides2024, Ianyoshi2025_variability, Zhang2025, Secunda2026}. In this scenario, super-Eddington accretion would lead to a puffed-up accretion disk where electron trapping could occur, leading to a dampening of any variability signal \citep{Ianyoshi2025_variability}. Additionally, studies at low redshift have found that the variability amplitude of AGN decreases with an increasing Eddington ratio at fixed bolometric luminosities \citep{MacLeod2010, Zhu2012}. The blue-shifted Balmer absorption commonly seen in LRDs \citep[e.g.,][]{Maiolino2024a, Taylor2024, Mathee2023,Matthee2026, Davis2026} can be fueled by super-Eddington outflows \citep{Naidu2026} and super-Eddington accretion can also help explain the lack of observed X-ray emission in LRDs \citep[e.g.,][]{Pacucci2024, Madau2024, Lambrides2026_superedd}.

The (marginal) detection of broad \Ha\ variability in LRDs OCEANS-100424 and the GlimmIr suggests that the broad \Ha\ is being emitted near the central engine of the LRD. The observed broad \Ha\ variability indicates a direct sight line to the accretion disk and/or BLR of the AGN without significant reprocessing or scattering. 
Observing variability in two LRDs implies a BLR that is similar to local quasars, in terms of compact size (i.e., proximity to a central BH) and clear line of sight \citep[e.g.,][]{VandenBerk2004}. Single-epoch BH mass estimators derived from the local Universe are constructed from RM experiments where virial masses of AGN are measured \citep[e.g.,][]{Greene2005, ReinesVolonteri2015, DallaBonta2025}.  RM works by measuring the lag in variability between the AGN continuum and the variability in broad permitted lines (e.g., \Ha, \Hb), which yields estimates of the BLR radius ($R_{BLR}$) enabling measurements of virial masses \citep{Blandford1982}. The detection of broad \Ha\ variability in LRDs supports the assumption that the broad Balmer emission is coming from a compact, virialized BLR close to the BH. This in turn provides confidence that the assumptions underlying the single-epoch BH mass estimators are potentially applicable to LRDs. However, even if both LRDs and quasars have virialized BLRs, there could still be significant differences in BLR kinematics and geometry in LRDs compared to local quasars that lead to differences in the single-epoch mass estimators.

Additionally, the (marginal) detection of BL variability in at least some LRDs provides evidence against a purely scattering-dominated origin \citep[e.g.,][]{Rusakov2026} for the broad \Ha\ profiles in this population. In the scattering-dominated emission model, any variability signal is reprocessed and smoothed. For OCEANS-100424 and the GlimmIr, \citet{Davis2026} finds that their broad \Ha\ profiles are well described with a single Gaussian, and show no evidence for the exponential wings expected from scattering-dominated line profiles. In these two sources, the combination of the Gaussian profile shape with the detected variability supports a virialized BLR origin consistent with local quasars.

The lack of variability in the other 4 LRDs does not necessarily mean that they are nonvariable, and might just indicate lack of significant flux changes over two observations epochs. We next investigate the statistical variability characteristics of our LRD sample and compare to the variability of low-redshift quasars.

\subsection{Do LRDs have similar variability to SDSS quasars?}

\begin{figure*}
    \centering
     \includegraphics[width=.8\linewidth]{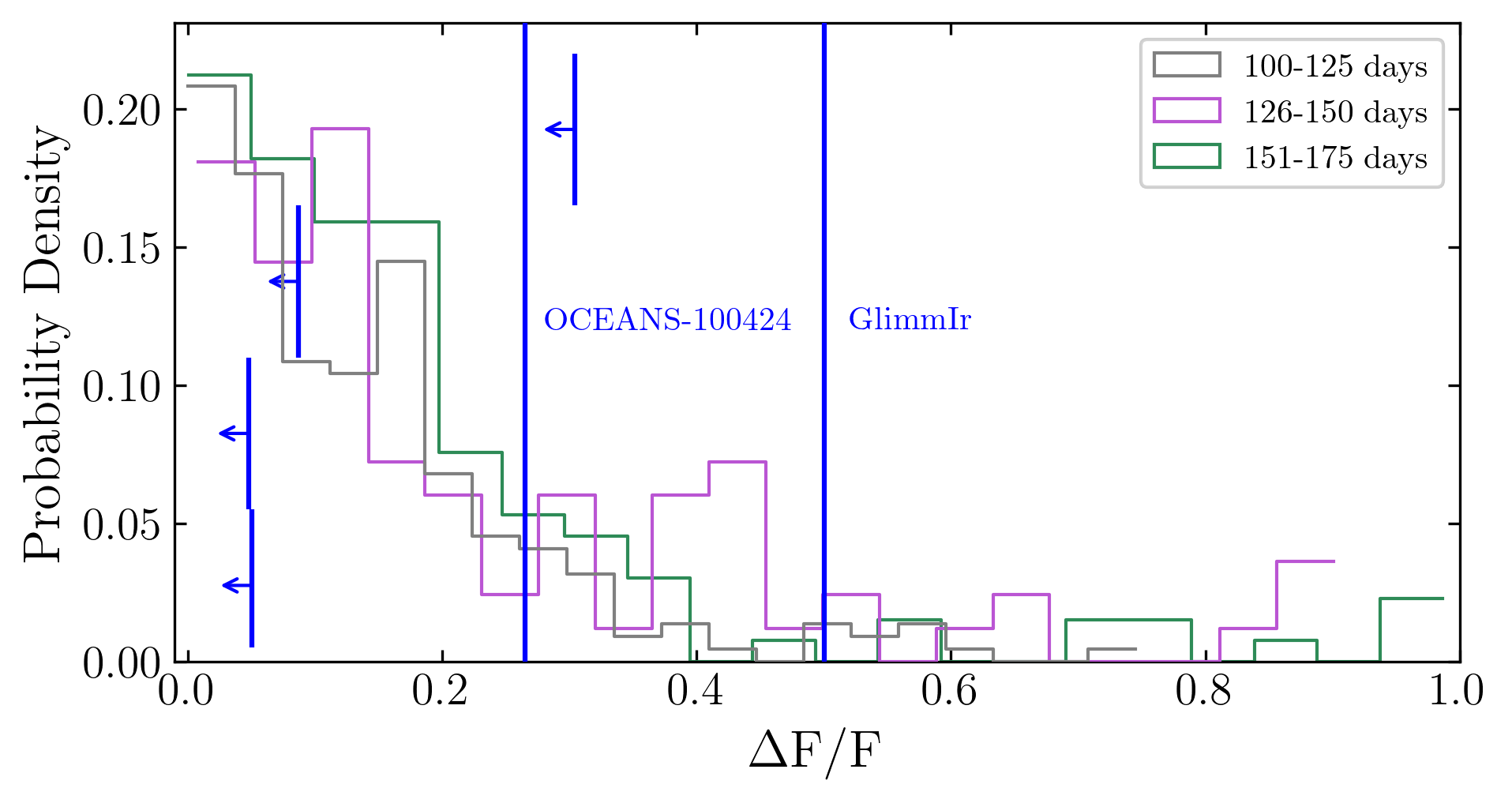}
     \caption{Probability distribution of $\Delta \rm F/F$ for a sample of SDSS-RM quasars. The SDSS-RM sample is divided into three rest-frame time bins: 100-125 days (gray), 126-150 days (pink), and 151-175 days (green). The two solid blue lines denote the $\Delta \rm F/F$ measurements for OCEANS-100424 and OCEANS-35829. Upper limits on $\Delta \rm F/F$ for the 4 other LRDs in this study are shown with the blue arrows.}\label{fig: sdss comparison}
\end{figure*}

To test if our results are consistent with the broad \Ha\ variability observed in typical AGN, we construct a comparison sample of AGN by utilizing spectroscopic light curves from the SDSS-RM survey \citep{Shen2015a, Shen2019b, Shen2024}. We select quasars from the SDSS-RM survey that have previously measured \Ha\ reverberation lags (23 sources) and identify all pairs of spectroscopic epochs separated by 100-175 rest-frame days. For each source, we draw 20 random spectroscopic pairs that satisfy the observational baseline criterion. For each epoch pair, we then fit the $\Ha + \NII$ complex with the two-component narrow+broad Gaussian model described in \S \ref{sec: emission line fitting}. Figure \ref{fig: sdss comparison} shows the distribution of fractional flux variability ($\Delta \rm F/F$) for the SDSS-RM quasar sample in three rest-frame time bins: 100-125 days (gray), 126-150 days (pink), and 151-175 days (green). The two solid blue lines denote the measured $\Delta \rm F/F$ for OCEANS-100424 and the GlimmIr, while the solid blue arrows show the $\Delta \rm F/F$ upper limits for the non-variable LRDs.

We then quantify the significance of the observed broad \Ha\ variability in OCEANS-100424 and the GlimmIr, by comparing their measured $\Delta \rm F/F$ values against the typical AGN $\Delta \rm F/F$ distribution constructed from the SDSS-RM sample. We generate $10^5$ Monte Carlo realizations by randomly drawing six quasars, without replacement, from the SDSS-RM distribution, that matches our LRD sample size. For each Monte Carlo realization, we determine whether the two most variable quasars exhibit $\Delta \rm F/F$ values at least as large as measured for OCEANS-100424 and the GlimmIr. We find that $20 \%$ of the random SDSS-RM realizations contain two quasars with $\Delta \rm F/F$ values at least as large as OCEANS-100424 and the GlimmIr. The observed variability we find in OCEANS-100424 and the GlimmIr is broadly ($ <2\sigma$) consistent with SDSS quasar variability. 

To determine if the GlimmIr is driving the low probability from above, we determine whether the two most variable quasars in the Monte Carlo realizations exhibit $\Delta \rm F/F$ values similar (within $1\sigma$) to OCEANS-100424. We find that $86\%$ of the random SDSS quasar realizations contain two quasars that have $\Delta \rm F/F$ values consistent with OCEANS-100424.

We also determine how often four of the SDSS quasars exhibit variability below the detection threshold of the nonvariable LRDs. We find that $30 \%$ of the random SDSS-RM realizations contain four quasars below this detection threshold. This occurrence rate is partially driven by OCEANS-161695 having a large $1\sigma$ upper limit ($\Delta \rm F/F \sim 0.30$) resulting from lower $\text{SNR}$ broad \Ha\ flux measurements. The other three nonvariable LRDs have $0.05 <\Delta \rm F/F  = 0.09$. With higher SNR observations of OCEANS-161695, the upper limit on its variability would likely be more stringent, leading to a stronger constraint on the fraction of SDSS-RM realizations consistent with the nonvariable LRD sample.



To model the full LRD sample simultaneously, we again generate $10^5$ Monte Carlo realizations by randomly drawing six quasars from the SDSS-RM distribution and determine when two quasars are consistently variable with OCEANS-100424 and the GlimmIr, while the other 4 quasars are below the detection threshold of the nonvariable LRDs. We find that the probability of both of these conditions being met is $4.71\%$, corresponding to a $\sim 2\sigma$ departure from typical quasar variability. 

The observed variability of 2 of our sources is broadly ($<2\sigma$) consistent with SDSS quasar variability. The primary outlier is the GlimmIr; which was intentionally pre-selected for follow-up observations in the OCEANS survey due to known variability \citep{Lambrides2026}. We find it difficult to reproduce 4 nonvariable sources and one extremely variable source from SDSS quasars. For a more detailed analysis on the potential drivers of the extreme variability seen in the GlimmIr, see Lambrides+2026b in prep.


\subsection{Comparison to Previous Work}

The recent TWINKLE survey (GO-7404, PIs: R. P. Naidu, J. Matthee, and J. Chisholm) conducted a NIRCam slitless spectroscopy program designed to probe LRD variability (see \citealt{Liu2026} for a full description of the program). \citealt{Liu2026} surveyed 18 LRDs at $z = 3.9-6.8$ with rest-frame observation baselines of $\sim 140-220$ days. In their study, they detect no variability in the photometry, \Ha\ line flux, or line shape across their sample.

For the average broad \Ha\ luminosity of OCEANS-100424, the $3\sigma$ sensitivity limit of $\Delta \rm F/F$ for the NIRCam grism is $21\%$. OCEANS-100424 sits right on the detection limit of the grism and may not have been robustly detected in the TWINKLE survey. This suggests that the discrepancy between the OCEANS-100424 detection and the lack of variability seen in the TWINKLE survey may be attributed to survey/instrumental resolution. 

The GlimmIr, if observed in the TWINKLE survey, would have been robustly detected at $2\times$ the survey's $3\sigma$ sensitivity limit. The GlimmIr is especially difficult to explain because it is more variable than typical SDSS quasars. The extreme variability of this source is described further in Lambrides+2026b in prep.  

Our sample exhibits no significant detection of continuum variability across all 6 LRDs. This could suggest that the continuum is not tracing the accretion disk emission or the continuum emission is reprocessed. We also note that the lack of continuum variability could be due to the weak continuum detection in the $R \sim 2700$ grating data. \citet{Stone2026} analyzed a sample of LRDs and BLAGN observed over rest-frame timescales of 1-3 months using the lower-resolution PRISM ($R \sim 100-300$) and NIRCam data from the North ecliptic pole EXtragalactic Unified Survey (NEXUS; \citealt{Shen2024_nexus}). \citet{Stone2026} found a lack of  ensemble continuum variability in their LRD/BLAGN sample ($\lesssim 3 \%$). The consistency between our non-detection of continuum variability and the \citet{Stone2026} result might suggest that the lack of continuum variability seen here is an intrinsic LRD property rather than an artifact of our lower-SNR grating measurements.





\subsection{The diversity of LRDs}
Recent works have suggested a diversity of LRDs rather than one homogeneous population \citep[e.g.,][]{deGraaff2025b, Barro2025, Perez-Gonzalez2026, Billand2026}. Our results suggest that LRDs may exhibit a diversity of variability properties tied to different evolutionary phases. OCEANS-100424 and GlimmIr, which show evidence for broad \Ha\ flux variability, may represent a post ``blow-out" phase in which the covering-factor of circum-nuclear gas begins to decrease, resulting in partial sight-lines to the BLR \citep{Lambrides2026}. A declining covering fraction could also reduce the prominence of the exponential BL wings seen in the \Ha\ emission line profile as less emission is reprocessed by scattering in a dense gas envelope \citep{Rusakov2026}.

Interestingly, \citet{Barro2025} classifies OCEANS-100424 and the GlimmIr (and also OCEANS-20504) into the same LRD subtype, defined by red optical colors and strong Balmer break strengths. Large Balmer breaks are normally associated with substantial covering fractions, which is hard to reconcile with a post blow-out evolutionary scenario. One way to address this discrepancy would be to invoke a clumpy medium  which allows intermittent, variable sight-lines to the BLR while still maintaining a large global Balmer break strength \citep{Tang2026, Ji2026_fuv} or a dynamically driven variability scenario. The physical mechanisms that drive the extreme variability of the GlimmIr will be discussed in more detail in Lambrides+2026b in prep.

\section{Conclusions}\label{sec: conclusions}
The unexpected population of LRDs discovered with JWST has challenged our current understanding of BH-galaxy co-evolution in the early Universe \citep[e.g,][]{Mathee2023, Kocevski2023}. Understanding the physical mechanisms that drive their observed properties remains one of the precedent questions in astrophysics. This work investigates the AGN nature of LRDs by searching for evidence of BL and continuum variability in a sample of six LRDs re-observed by the $R \sim 2700$ OCEANS spectroscopic survey.

 We find marginal evidence for broad \Ha\ variability in 2 LRDs, OCEANS-100424 ($2.1\sigma$) and the GlimmIr ($1.5\sigma$). 
The remaining four LRDs in our sample do not show evidence for BL variability. We also find no evidence for continuum variability across our sample. The marginal detection of broad \Ha\ variability suggests that the broad \Ha\ is being emitted near the central engine of the LRD, implying a direct line of sight to the accretion disk and/or BLR without any significant reprocessing or scattering.

To fully study the range of LRD variability, deep multi-epoch spectroscopy is needed. Resolving the frequency and timescale of LRD variability requires long baseline spectroscopic monitoring of these systems. Further expanding the sample of LRDs with repeat spectroscopy is necessary for studying their diversity and is critical for our comprehensive understanding of BH-galaxy co-evolution in the early Universe.


\section{Acknowledgments}
(Some of) The data products presented herein were retrieved from the Dawn JWST Archive (DJA). DJA is an initiative of the Cosmic Dawn Center (DAWN), which is funded by the Danish National Research Foundation under grant DNRF140. We thank the CEERS, CAPERS, THRILS, and RUBIES teams for their effort designing and executing their programs and for making the data publicly available.

MB acknowledges support from a NSF Graduate Research Fellowship award number 2136520. MB, KD, JRT, and RCS acknowledge support from NASA JWST-GO-08410.018. The JWST data presented in this article were obtained from the Mikulski Archive for Space Telescopes (MAST) at the Space Telescope Science Institute. The specific observations analyzed in this paper can be accessed via \dataset[doi: 10.17909/tj0h-cj44]{https://doi.org/10.17909/tj0h-cj44}.

\facility{\textit{JWST}}
\software{\texttt{astropy}: \cite{Astropy2013, Astropy2018, Astropy2022}, \texttt{scipy}: \cite{scipy2020}, \texttt{emcee}: \cite{emcee}, \texttt{numpy}: \cite{numpy}, \texttt{matplotlib}: \cite{matplotlib}}

\appendix

\begin{figure*}
    \centering{}
    \includegraphics[width = \linewidth]{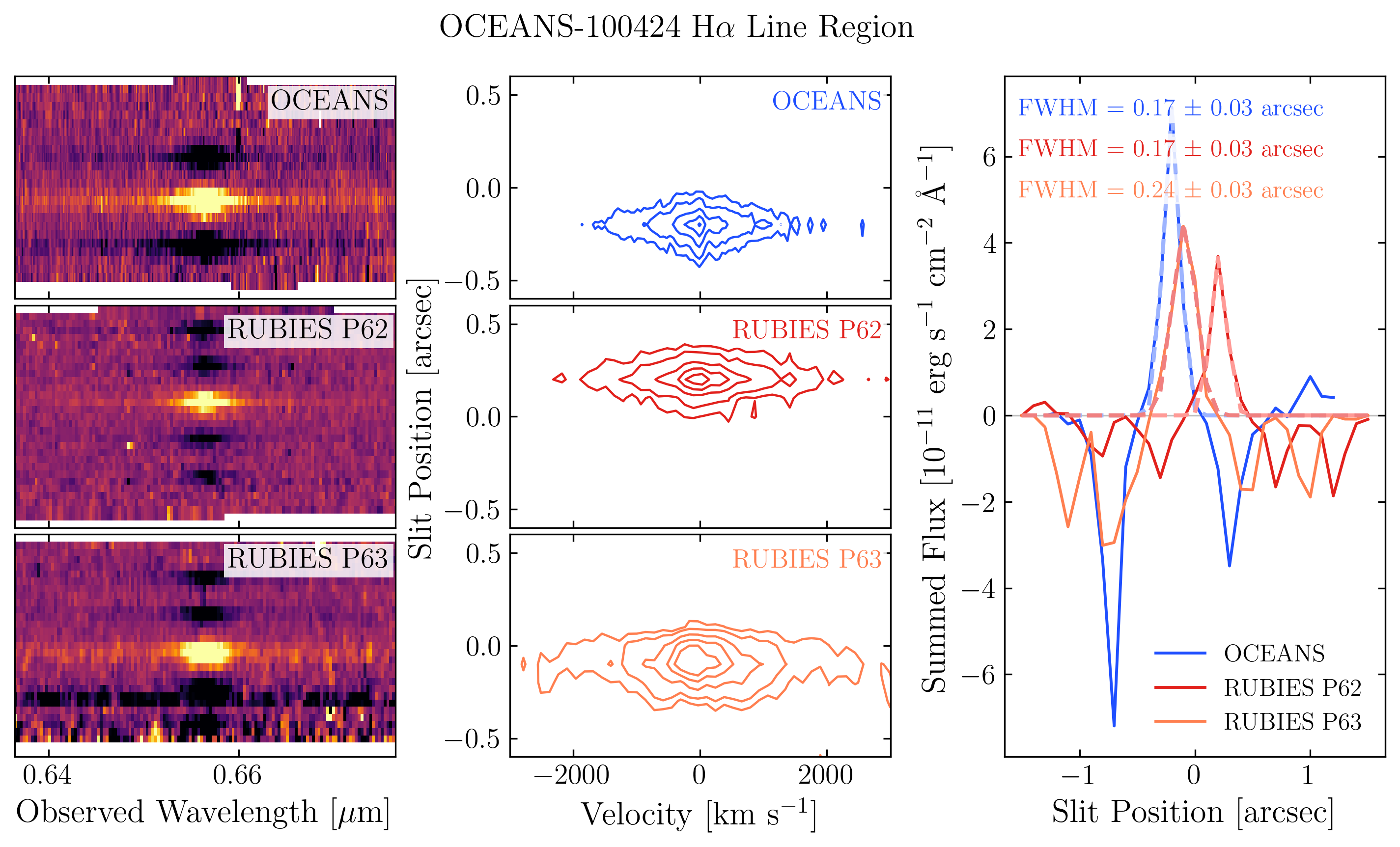}
    \caption{\textit{Left}: 2D spectra for each observation epoch of LRD OCEANS-100424. \textit{Middle:} 2D contours in velocity–slit-position space for the OCEANS epoch (top, blue), RUBIES P62 (middle, red), and RUBIES P63 (bottom, orange), centered on the broad \Ha\ emission line region. \textit{Right:} Spatial profile of the summed line flux as a function of slit position for each spectroscopic epoch with best-fit Gaussian models shown as dashed curves.}
    \label{fig: appendix fig halpha}
\end{figure*}

 As described in \S \ref{sec:Spectroscopic Data}, OCEANS-100424 was observed in 2 separate RUBIES pointings, P62 and P63. We initially noticed artifacts present in the 2D spectra of RUBIES P63 that are not present in the OCEANS or P62 2D spectra. These artifacts are apparent in the \Ha\ spectral region of RUBIES P63, where the bottom nod is noticeably different from the top nod, is shown in the bottom left panel of Figure \ref{fig: appendix fig halpha}. This nod-to-nod discrepancy is not seen in the OCEANS or RUBIES P62 observation of this source. The 2D contour plot of the \Ha\ spectral region in RUBIES P63, Figure \ref{fig: appendix fig halpha} bottom middle panel, also appears asymmetrical compared to the OCEANS and RUBIES P62 contours. 

 To evaluate the asymmetry seen in RUBIES P63, we construct a profile of the cross-dispersion (spatial) direction by summing the flux in the \Ha\ region for each row along the slit. The \Ha\ spatial profiles for the three different OCEANS-100424 epochs are shown in the right-hand panel of Figure \ref{fig: appendix fig halpha}. We fit a single Gaussian to the spatial profile and find a $\rm FWHM = 0.17 \pm 0.03$ arcsec for both the OCEANS and RUBIES P62 epoch. We find a $\rm FWHM = 0.24 \pm 0.03$ for the RUBIES P63 epoch. This implies a difference in the \Ha\ spatial profile of RUBIES P63 when compared to OCEANS and RUBIES P62.

 We elect to remove P63 from our analysis due to the artifacts present in the 2D spectra and the broadened \Ha\ spatial profile compared to the other two epochs of observation. The difference in spatial profile could result in a change to the \Ha\ flux falling in the slit relative to the other epochs and result in a bias to the slit-loss correction, which assumes a consistent source-profile across the different epochs.

\begin{figure*}
    \centering{}
    \includegraphics[width = \linewidth]{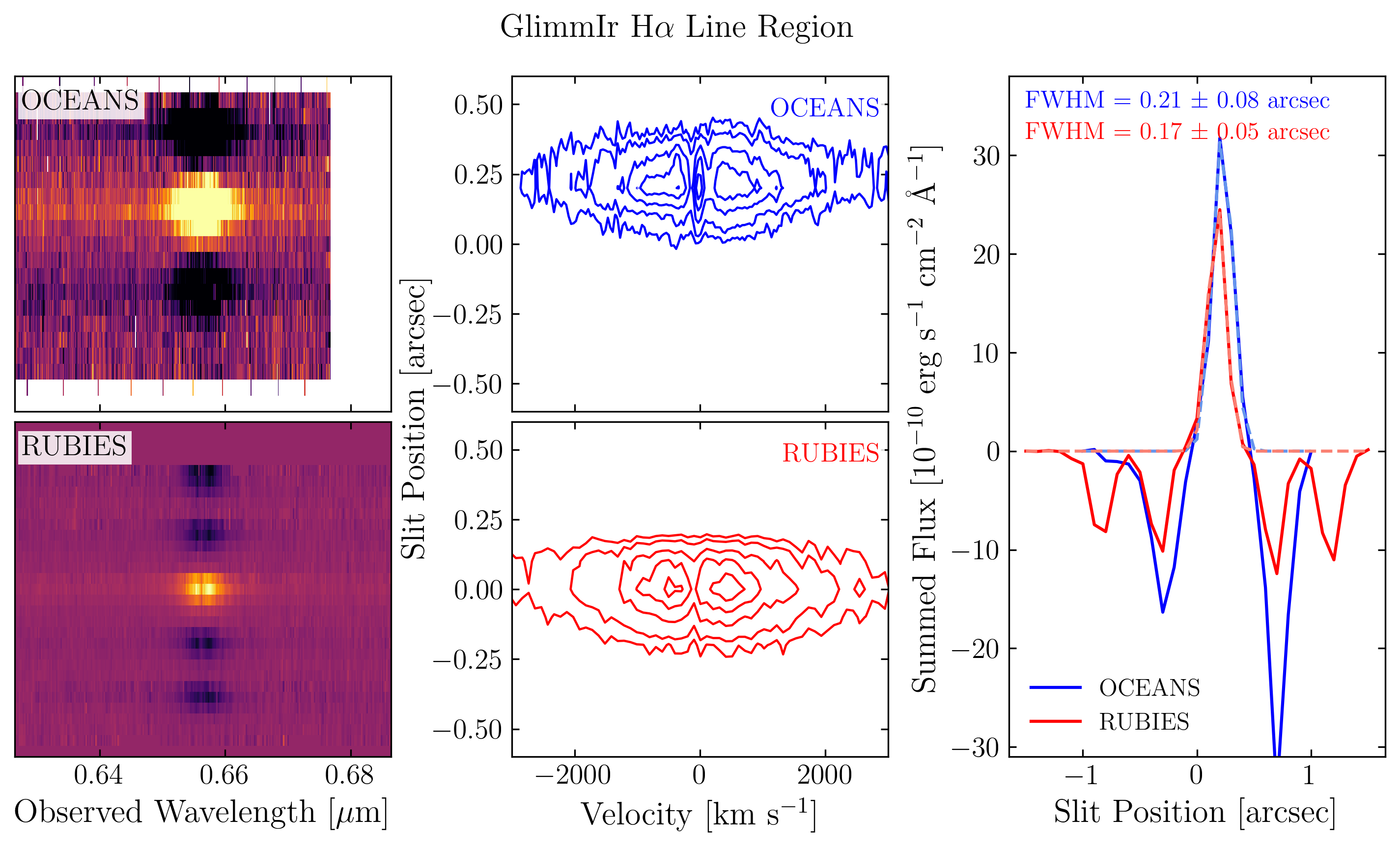}
    \caption{\textit{Left}: 2D spectra for each observation epoch of LRD the GlimmIr (OCEANS-35829). \textit{Middle:} 2D contours in velocity–slit-position space for the OCEANS epoch (top, blue) and RUBIES epoch (bottom, red) centered on the broad \Ha\ emission line region. \textit{Right:} Spatial profile of the summed line flux as a function of slit position for each spectroscopic epoch with best-fit Gaussian models shown as dashed curves}\label{fig: appendix glimmir}
\end{figure*}

 We additionally show the 2D spectra, 2D contours, and cross-dispersion spatial profile for the GlimmIr in Figure \ref{fig: appendix glimmir}. We find no apparent artifacts in the 2D spectra, no asymmetries in the 2D contours, and a consistent \Ha\ spatial extent.  
 
\clearpage
\bibliography{citations}

\end{document}